\documentclass[dvips]{article}
\newcommand{\doublespacing}{\let\CS=\@currsize\renewcommand{\baselinesstrech}
{2.0}\tiny\CS}

\begin{document}

\textwidth 16cm
\newcommand{\bd}{\begin{document}}
\newcommand{\ed}{\end{document}}
\newcommand{\bc}{\begin{center}}
\newcommand{\ec}{\end{center}}
\newcommand{\bfr}{\begin{flushright}}
\newcommand{\efr}{\end{flushright}}
\newcommand{\lt}{\left}
\newcommand{\rt}{\right}
\newcommand{\vs}{\vspace}
\newcommand{\hs}{\hspace}
\newcommand{\beq}{\begin{equation}}
\newcommand{\eeq}{\end{equation}}
\newcommand{\lb}{\linebreak}
\newcommand{\pb}{\pagebreak}
\newcommand{\mb}{\makebox}
\newcommand{\fb}{\framebox}
\newcommand{\mc}{\multicolumn}
\newcommand{\ben}{\begin{enumerate}}
\newcommand{\een}{\end{enumerate}}
\newcommand{\bit}{\begin{itemize}}
\newcommand{\eit}{\end{itemize}}
\newcommand{\ol}{\overline}
\newcommand{\un}{\underline}
\newcommand{\lefq}{\lefteqn}
\newcommand{\ba}{\begin{array}}
\newcommand{\ea}{\end{array}}
\newcommand{\beqa}{\begin{eqnarray}}
\newcommand{\eeqa}{\end{eqnarray}}
\newcommand{\beqas}{\begin{eqnarray*}}
\newcommand{\eeqas}{\end{eqnarray*}}
\newcommand{\bfg}{\begin{figure}}
\newcommand{\efg}{\end{figure}}
\newcommand{\bds}{\begin{displaymath}}
\newcommand{\eds}{\end{displaymath}}
\newcommand{\btb}{\begin{tabbing}}
\newcommand{\etb}{\end{tabbing}}
\newcommand{\para}{\parallel}
\newcommand{\pad}{\partial}
\newcommand{\nn}{\nonumber}
\newcommand{\la}{\leftarrow}
\newcommand{\ra}{\rightarrow}
\newcommand{\lgla}{\longleftarrow}
\newcommand{\lgra}{\longrightarrow}
\newcommand{\La}{\Leftarrow}\newcommand{\Ra}{\Rightarrow}
\newcommand{\Lra}{\Leftrightarrow}
\newcommand{\Lgla}{\Longleftarrow}
\newcommand{\Lgra}{\Longrightarrow}
\newcommand{\bm}{\boldmath}
\newcommand{\lan}{\langle}
\newcommand{\ran}{\rangle}
\renewcommand{\a}{\alpha}
\renewcommand{\b}{\beta}
\newcommand{\g}{\gamma}
\newcommand{\G}{\Gamma}
\renewcommand{\d}{\delta}
\newcommand{\eps}{\epsilon}
\newcommand{\Th}{\Theta}
\newcommand{\s}{\sigma}
\newcommand{\lam}{\lambda}
\newcommand{\D}{\Delta}
\newcommand{\vare}{\varepsilon}
\newcommand{\pr}{\prime}
\newcommand{\ro}{\rho}
\newcommand{\nab}{\nabla}
\newcommand{\m}{\mu}
\newcommand{\n}{\nu}
\newcommand{\Sg}{\Sigma}
\newcommand{\p}{\pi}
\newcommand{\R}{I\!\!R}
\newcommand{\om}{\omega}
\newcommand{\Om}{\Omega}
\newcommand{\ze}{\zeta}
\newcommand{\vart}{\vartheta}
\newcommand{\tri}{\triangle}
\newcommand{\f}{\frac}
\newcommand{\iny}{\infty}
\newcommand{\pro}{\propto}
%\input{fqhsc}
%\input{state}
%\input{acc}
%\input{fs}
%\ed
\bc {\huge Shape invariance approach to exact solutions of the Klein-Gordon equation} \ec

\vs{1cm}

\bc
{\it T. Jana{\footnote {e-mail : tapas$_{-}$r@isical.ac.in} and P. Roy{\footnote{e-mail : pinaki@isical.ac.in}}\\
Physics \& Applied Mathematics Unit \\
Indian Statistical Institute \\
Kolkata - 700 108, India.}} \ec
\vs{4.5cm}

\bc {\large {\un{Abstract}}} \ec 
Using the shape invariance property we obtain exact solutions of the $(1+1)$ dimensional Klein-Gordon equation for certain types of scalar and vector potentials. We also discuss the possibility of obtaining real energy spectrum with non-Hermitian interaction within this framework.\\
\pb

\section{Introduction}
In view of their importance in the study of various relativistic systems under the influence of strong potentials \cite{greiner},
there have been a growing interest in obtaining exact solutions of relativistic wave equations, particularly the Klein-Gordon (KG) equation \cite{chen,castro,al}. In many of these of papers exact solutions were obtained for $(1+1)$ dimensional or the $s$ wave KG equation for different choices of the energy independent vector and the scalar potential. Furthermore the KG equation has also been considered in the context of non-Hermitian/$\cal{PT}$ symmetric interactions \cite{simsek}. Here our objective is to expand the class of exactly solvable KG equation for real (and unequal) as well as non-Hermitian vector and scalar interactions and supplement earlier results \cite{chen}. Also, instead of directly solving the relevant differential equation, we shall use algebraic techniques, namely, shape invariance \cite{khare} and intertwinning property to obtain exact solutions.

\section{Real scalar and vector potentials}
To begin with let us briefly recall the notion of shape invariance \cite{khare}. We note that a supersymmetric pair of potentials $U_{\pm}(x;a_1)(=W^2(x;a_1)\pm W'(x;a_1))$ are called shape invariant if  
\beq
\displaystyle U_+(x;a_1) = U_-(x;a_2) + R(a_1)\label{si}
\eeq
where $a_1$ is a set of parameters, $a_2$ is a function of $a_1$ and $R(a_1)$ is independent of $x$. It can be shown that for such potentials the energy spectrum corresponding to $U_-(x;a_1)$ is given by
\beq
\displaystyle E_{n}^- = \sum_{k=1}^n R(a_k)\label{energy}
\eeq
while the corresponding wave function is given by
\beq
\displaystyle \psi_n^-(x;a_1)\propto A^{\dagger}(x;a_1)A^{\dagger}(x;a_2)....A^{\dagger}(x;a_n)\psi_0^-(x;a_{n+1})\label{wf}
\eeq
where $\displaystyle{A(x;a_1)=-\f{d}{dx}+W(x;a_1)}$ and $\psi_0^-(x;a_{n+1})$ is the ground state wave function corresponding to $U_-(x;a_{n+1})$.

We shall now use the above results to obtain solutions of Klein-Gordon equation. Let us now consider stationary Klein-Gordon equation of the form
\beq
\displaystyle -\f{d^2\psi}{dx^2} + [(m+S(x))^2-(E-V(x))^2]\psi = 0 \label{kg1}
\eeq
where $V(x)$ and $S(x)$ denote the vector and the scalar potentials respectively. Written in full, Eq.(\ref{kg1}) becomes
\beq
\displaystyle -\f{d^2\psi}{dx^2} + U(x)\psi = \eps \psi \label{kg2}
\eeq
where
\beq
\displaystyle U(x) = S^2(x)-V^2(x)+2[mS(x)+EV(x)]~~,~~\eps = E^2-m^2 \label{u}
\eeq 
To use the concept of shape invariance it is now necessary to identify (\ref{u}) with one of $U_{\pm}$. To do this we choose the vector and scalar potentials to be of the form
\beq
\displaystyle V(x) = V_0 f(x)~~,~~S(x) = S_0 f(x)
\eeq
Then from (\ref{u}) it follows that 
\beq
\displaystyle U(x) = (S_0^2-V_0^2)f^2(x) + 2(mS_0+EV_0)f(x)\label{u10}
\eeq
Thus to put (\ref{u10}) in the form $U_{\pm}$ it is necessary to choose $f(x)$ such that $f'(x)$ can be expressed in terms of $f(x)$ and also impose suitable constraints on the coupling constants. Furthermore one has to be careful so that $V(x),S(x)$ do not become energy dependent. We shall now obtain exact solutions of (\ref{kg2}) for different choices of $f(x)$. 
\vspace{.25cm}

{\underline{\bf Case 1.}} 
We first consider 
\beq f(x)=tanh x
\eeq
In this case Eq.(\ref{kg2}) can be written as
\beq
\displaystyle -\f{d^2\psi}{dx^2} + U_1(x)\psi = \eps_1 \psi \label{kg3}
\eeq
where
\beq
\displaystyle U_1(x) = -(S_0^2-V_0^2)sech^2x + 2(mS_0+EV_0)tanhx~~,~~\eps_1 = E^2-m^2-S_0^2+V_0^2 \label{u1}
\eeq 
It is readily seen that if  $S_0^2>V_0^2$ then $U_1(x)$ can be treated as a shape invariant potential admitting discrete eigenvalues \cite{khare}. In this case we have
\beq
\ba{lcl}
\displaystyle{W(x;a_1) = A tanhx+\f{B}{A}~,~~ a_i = (A-i+1)},\\\\
\displaystyle R(a_i) = [(A-i+1)^2-(A-i+2)^2] + B^2\left[\f{1}{(A-i+1)^2}-\f{1}{(A-i+2)^2}\right]
\ea
\eeq
where
\beq
\displaystyle A = \f{-1+\sqrt{1+4(S_0^2-V_0^2)}}{2}~,~B = mS_0+E_nV_0 \label{ab}
\eeq 
Then from (\ref{energy}) it follows that
\beq
\displaystyle \eps_{1n} = \sum_{i=1}^n R(a_i) = -(A-n)^2 -  \f{B^2}{(A-n)^2}~,~n=0,1,2,...<(A-\sqrt{|B|})\label{eps1}
\eeq
We note that in comparison to nonrelativistic Schr\"odinger equation, in the present case the number of discrete energy levels depends on the energy $E_n$ rather than on $n$. Now from (\ref{eps1}) and (\ref{ab}) the energy eigenvalues are found to be 
\beq
\displaystyle E_n^\pm = \f{-Q\pm \sqrt{Q^2-4P_nR_n}}{2P_n}\label{en1}
\eeq
where
\beq
\ba{lcl}
P_n &=&\displaystyle (A-n)^2 + V_0^2\\
Q &=&\displaystyle 2mS_0V_0\\
R_n &=&\displaystyle (A-n)^4+m^2S_0^2-(A-n)^2(m^2+S_0^2-V_0^2)\\
\ea
\eeq
Now using (\ref{wf}) and noting that the ground state wave function can be obtained from the relation $A(x;a_1)\psi_0(x;a_1)=0$, the wave functions corresponding to (\ref{en1}) can be found to be \cite{khare,nieto}
\beq
\displaystyle \psi_n(x;a_1)\propto (1-tanh x)^{s_1^{\pm}/2}(1+tanh x)^{s_2^{\pm}/2}P_n^{(s_1^{\pm},s_2^{\pm})}(tanh x)\label{wf1}
\eeq
where the$(\pm)$ sign correspond to positive and negative energy,~ $\displaystyle{s_1^{\pm}=A-n+\f{B}{A-n},~ s_2^{\pm}=A-n-\f{B}{A-n}}$ and $P_n^{(a,b)}$ denotes Jacobi polynomials.
\vspace{.2cm}

It is important to note that not all the energy levels in (\ref{en1}) are acceptable. To determine the acceptable levels we need to examine the behaviour of the wave functions. From (\ref{wf1}) it is seen that the wave functions are acceptable if $s_1^{\pm},s_2^{\pm}>0$. But both $s_1^{\pm}$ and $s_2^{\pm}$ depend on $A$ and $B$ with $B$ depending on $E_n^\pm$. However the resulting dependence of $s_1^{\pm}$ and $s_2^{\pm}$ on $E_n^\pm$ is quite complicated and the above conditions has to be checked numerically for each level. We have computed the energy values and the parameters $s_{1,2}^\pm$ for different values of the input parameters $m,S_0,V_0$ and the results are given in tables 1 and 2. 
 
\vspace{.25cm}

{\underline{\bf Case 2.}} We shall now choose 
\beq f(x)=-e^{-x}\label{f2}
\eeq
This case was treated in ref\cite{chen} by solving the KG equation. Here we rederive the results via the shape invariance property and examine the acceptable energy levels. For the choice (\ref{f2}) we obtain from (\ref{u}) 
\beq
\displaystyle U_2(x) = (S_0^2-V_0^2)e^{-2x}-2(mS_0+EV_0)e^{-x}
\eeq
In this case it can be shown that  
\beq
\ba{lcl}
W(x;a_1) = A-Be^{-x}~,~~a_i=(A-i+1),\\\\
\displaystyle R(a_i) = (A-i+1)^2 - (A-i)^2
\ea
\eeq
where 
\beq
\displaystyle A^{\pm} = \f{mS_0+EV_0}{\sqrt{S_0^2-V_0^2}}-\f{1}{2}~,~~B = \sqrt{S_0^2-V_0^2}
\eeq
Thus we have 
\beq
\displaystyle \eps_{2n} = \sum_{i=1}^n R(a_i) = -(A-n)^2~,~~n=0,1,2,...<A
\eeq
and consequently
\beq
\displaystyle E_n^\pm = \f{-Q_n\pm\sqrt{Q_n^2-4PR_n}}{2P}\label{en2}
\eeq
where 
\beq
\ba{lcl}
P &=&\displaystyle S_0^2\\
Q_n &=&\displaystyle -2(n+\f{1}{2})V_0\sqrt{S_0^2-V_0^2}+2mV_0S_0\\
R_n &=&\displaystyle (n+\f{1}{2})^2(S_0^2-V_0^2)+m^2V_0^2-2(n+\f{1}{2})mS_0\sqrt{S_0^2-V_0^2}\\
\ea
\eeq
The wave functions corresponding to (\ref{en2}) are given by \cite{khare,nieto1}
\beq
\displaystyle \psi_n(x)\propto (2Be^{-x})^{A^\pm-n}e^{-Be^{-x}}L_{n}^{2A^\pm-2n}(2Be^{-x})\label{wf2}
\eeq
where $L_n^\alpha(x)$ denotes generalised Laguerre polynomials. From (\ref{wf2}) it follows that normalisability of the wave functions requires $B>0$ and $A^\pm >0$ ($A^\pm$ denotes the values of $A$ corresponding to positive and negative energy). The first of these conditions can be ensured by choosing $S_0^2>V_0^2$ and since $A^\pm$ depends on $E_n^\pm$, the second condition has to be checked for each individual quantum number $n$. Furthermore, the number of discrete energy states depends on the values of $A^\pm$. We have numerically evaluated the energy values and the parameter $A^\pm$ for some value of the input parameters $m,S_0$ and $V_0$ and the results are presented in table 3. 
\vspace{.25cm}

{\underline{\bf Case 3.}} Finally we consider 
\beq f(x)=x/2
\eeq
In this case we obtain
\beq
\displaystyle U_3(x) = \f{1}{4}(S_0^2-V_0^2)x^2+(mS_0+EV_0)x\label{u3}
\eeq 
Clearly $U_3(x)$ is a shape invariant potential with
\beq
W(x;a_1) = Ax + B~~,~~
\displaystyle a_i = A~~,~~R(a_i) = 2A
\eeq
where 
\beq
\displaystyle A^2 = \f{S_0^2-V_0^2}{4}~~,~~B = \f{mS_0+EV_0}{\sqrt{S_0^2-V_0^2}}
\eeq
Thus we get
\beq
\displaystyle \eps_{3n} = (2n+1)A-B^2
\eeq
so that 
\beq
\displaystyle E_n^\pm = \f{-Q_n\pm\sqrt{Q_n^2+4PR_n}}{2P_n}\label{en3}
\eeq
where
\beq
\ba{lcl}
P &=&\displaystyle S_0^2\\
Q &=&\displaystyle 2mS_0V_0\\
R_n &=&\displaystyle (S_0^2-V_0^2)^{3/2}(n+\f{1}{2})-m^2V_0^2
\ea
\eeq
The corresponding wave functions are given by
\beq
\displaystyle \psi_n(x)\propto e^{-y^2/2}H_n(y)~~,~~y = \sqrt{A}(x+\f{B}{A})\label{wf3}
\eeq
It follows from (\ref{u3}) that for discrete energy states the condition $S_0^2>V_0^2$ should hold and in this case all the energy levels exist. In table 4 we present some energy values for certain values of the parameters $m,V_0$ and $S_0$.

\section{Non-Hermitian scalar and vector potentials.} 
Here we shall consider non-Hermitian vector and scalar potentials yielding real energy. Generally there are two ways to generate non-Hermitian interaction. The first one involves considering complex coupling constants \cite{bender} while in the second approach one considers a complex coordinate shift \cite{znojil1}. Although the spectrum for potentials complexified via the first method can be obtained easily for Schr\"odinger equation, it requires extensive numerical computation in the case of KG equation. Consequently here we shall follow the second approach which is considerably simpler. Thus we consider vector and scalar potentials to be of the form
\beq
\displaystyle V(x) = V_0 f(x-ic) ~~,~~S(x) = S_0 f(x-ic)
\eeq
where $c$ is a real constant. As an illustration let us consider $f(x)=tanh(x)$. So from (\ref{u1}) we get
\beq
\displaystyle U_1^{NH}(x-ic) = -(S_0^2-V_0^2)sech^2(x-ic) + 2(mS_0+EV_0)tanh(x-ic)~~,~~\eps_1 = E^2-m^2-S_0^2+V_0^2 \label{u11}
\eeq
Comparing with (\ref{u1}) it is easy to see that the non-Hermitian potential $U_1^{nh}(x)$ is shape invariant. Also the effective coupling constants do not contain any imaginary term and since the shape invariance condition (\ref{si}) is coordinate independent, the (real) energy spectrum is given by (\ref{eps1}). 

Next to examine the behaviour of the system with respect to $\cal{PT}$ symmetry, we note that a Hamiltonian $H$ is said to be $\cal{PT}$ invariant if \cite{bender}
\beq
{\cal{PT}}H = H{\cal{PT}}
\eeq
where the space inversion ($\cal{P}$) and time reversal ($\cal{T}$) operators are defined by their actions on position, momentum and identity operators as \cite{bender}
\beq
\displaystyle {\cal{P}}x{\cal{P}} = -x~,~{\cal{P}}p{\cal{P}} = {\cal{T}}p{\cal{T}} = -p~,~{\cal{T}}(i.1){\cal{T}} = -i.1\\
\label{PT}
\eeq
Now using (\ref{PT}) it is easily seen that the non-Hermitian potential (\ref{u11}) is not $\cal{PT}$ symmetric i.e, $U_1^{NH}(x)\neq U_1^{NH*}(-x)$. Thus (\ref{u11}) is a non-Hermitian and non $\cal{PT}$ symmetric potential which has real spectrum. Clearly the complex coordinate shift method can be extended to the other cases considered here to obtain complex non $\cal{PT}$ symmetric potentials with real spectrum.

Finally, we explain the equivalence of a Hermitian and non-Hermitian system in a different way. First let us consider the operator \cite{ahmed}
\beq
\eta = e^{cp},~~c~ real
\eeq
with the properties
\beq
\eta V(x)\eta^{-1} = V(x-ic)~~,~~\eta p\eta^{-1} = p~~,~~\eta \psi(x) = \psi(x-ic)\label{property1}
\eeq
Then it follows that
\beq
\eta U_{1}(x)\eta^{-1} = U_1^{NH}(x-ic) \label{property2}
\eeq
Then using (\ref{property1}) and (\ref{property2}) it follows that if $E$ is an eigenvalue of the potential $U_1(x)$ with eigenfunction $\psi(x)$, then $E$ is also an eigenfunction of the potential $U_1^{NH}(x-ic)$ with eigenfunction $\eta \psi(x) = \psi(x-ic)$. Thus the two Hamiltonians are intertwined and share the same real spectrum.

\section{Conclusion}
Here we have obtained exact solutions for KG equation for a class of (energy independent) vector and scalar potentials using the shape invariance technique. It has also been shown that the KG equation admits real eigenvalues in the presence of non Hermitian vector and scalar potentials. It may also be noted that here we have considered shape invariance based on translation of parameters and it would be interesting to examine the possibility of obtaining exact solutions of KG equation for self similar potentials.

\pb
\begin{center}  

\begin{tabular}{|   l|   l|  l|    l |     l|   l   |  l  | l |   }
\hline $n$    &$E^{+}_{n}$    & $s^{+}_{1}$    & $s^{+}_{2}$& $E^{-}_{n}$ &$s^{-}_{1}$  & $s^{-}_{2}$      \\
\hline  $0$   &  $1.83314$    & $3.98281$    & $3.049$     & $-1.88921$     & $3.61226$    & $3.41955$      \\
\hline    $1$   &  $2.99136$    & $3.32952$    & $1.70229$     & $-3.09985$     & $2.48214$    & $2.54967$      \\
\hline $2$     &  $3.39932$    & $2.96043$    & $0.071382$     & $-3.68852$     & $1.32395$    & $1.70786$      \\
\hline 
\end{tabular}

\noindent
{\bf Table 1.}~~$s_1^{\pm}, s_2^{\pm}$ ~and~$E_n^{\pm}$~for~ $m=0.25$,~~ $S_{0}=4$~ and~ $V_{0}=0.35$ 
\label{tab:1}
\end{center}

\vspace*{.5cm}

\begin{center}

\begin{tabular}{|  l|  l|    l|    l |     l|   l   |  l  |  }
\hline $n$   &$E^{+}_{n}$    & $s^{+}_{1}$    & $s^{+}_{2}$& $E^{-}_{n}$ &$s^{-}_{1}$  & $s^{-}_{2}$      \\
\hline  $0$    &  $1.791$    & $4.26304$    & $2.76877$     & $-1.90315$     & $3.8953$    & $3.13652$      \\
\hline    $1$    &  $2.8921$    & $3.71318$    & $1.31863$     & $-3.10908$     & $2.87833$    & $2.15348$      \\
\hline 
\end{tabular}

\noindent
{\bf Table 2.}~~$s_1^{\pm} ,s_2^{\pm} ~and~E_n^{\pm}~for~m=0.5,~~ S_{0}=4~$ and ~ $V_{0}=0.35$
\label{tab:2}
\end{center}

\vspace*{.5cm}

\noindent
\begin{center}
\begin{tabular}{|  l|    l|    l |     l|   l   |   }
\hline $n$    &$~~E^{+}_{n}~~$    & $~~A^{+}~~$   & $~~E^{-}_{n}~~$ &$~~A^{-}~~$     \\
\hline  $~~0~~$    &   $~~1.08989~~$    & $~~1.17139~~$     & $~~-1.22751~~$    & $~~1.02626~~$      \\
\hline  $~~1~~$    &  $~~1.58713~~$    & $~~1.20252~~$    & $~~-1.6~~$    & $~~1.00294~~$      \\
\hline
\end{tabular}

\noindent
{\bf Table 3.}~~$E_n^{\pm}~and~A^{\pm}~for~m=1.6,~~ S_{0}=4~ and~ ~V_{0}=0.25$
\label{tab:3}
\end{center}

\vspace*{.5cm}
\noindent

\begin{center}
\begin{tabular}{|  l|    l|    l |      }
\hline $~~n~~$    &$~~E^{+}_{n}~~$      & $~~E^{-}_{n}~~$     \\
\hline  $~~0~~$    &   $~~1.36234~~$     & $~~-1.44984~~$          \\
\hline  $~~1~~$    &   $~~2.39166~~$     & $~~-2.47916~~$          \\
\hline  $~~2~~$    &   $~~3.10035~~$     & $~~-3.18785~~$          \\
\hline
\end{tabular}

\noindent
{{\bf Table 4.}~~$E_n^\pm~for~m=0.5,~~ S_{0}=4~ $\\
$and~ V_{0}=0.35$}
\ec

\ed
\begin{thebibliography}{99}
\bibitem{greiner}W. Greiner, Relativistic Quantum Mechanics, 3rd Edition, Springer-Verlag, Berlin 2000.
\bibitem{chen} G. Chen, Z. Chen and Z. Lou, Phys.Lett {\bf A331}, (2004) 374\\
G. Chen, Phys.Lett {\bf A339}, (2005) 300\\
G. Chen, Z. Chen and P. Xuan, Phys.Lett {\bf A352}, (2006) 317\\
A. de Souza Dutra and G. Chen, Phys.Lett {\bf A349}, (2006) 297.
\bibitem{castro} A. S. de Castro, Phys.Lett {\bf A338}, (2005) 81\\
L. Z. Yi, Y. F. Diao, J. Y. Liu and C. S. Jia, Phys.Lett {\bf A333}, (2004) 212\\
Y. F. Diao, L. Z. Yi and C. S. Jia, Phys.Lett {\bf A332}, (2004) 157.
\bibitem{al} A. D. Alhaidari, H. Bahlouli and A. Al-Hasan, Phys.Lett {\bf A349}, (2006) 47.
\bibitem{simsek} M. Simsek and H. Egrifes, J.Phys {\bf A37}, (2004) 4379\\
M. Znojil, Czech J.Phys {\bf 54}, (2004) 1143. 
\bibitem{khare} F. Cooper, A. Khare and U. Sukhatme, Supersymmetry in Quantum Mechanics, (World Scientific, 2001).
\bibitem{nieto} M.M. Nieto, Phys.Rev {\bf A17}, (1978) 1273.
\bibitem{nieto1} M.M. Nieto and L.M. Simmons, Phys.Rev {\bf A19}, (1979) 438.
\bibitem{bender} C.M. Bender and S. Boettcher, Phys.Rev.Lett {\bf 80}, (1998) 5243.
\bibitem{znojil1} M. Znojil, Phys.Lett {\bf A259}, (1999) 220.
\bibitem{ahmed} Z. Ahmed, Phys.Lett {\bf A290}, (2001) 19.
\end{thebibliography}
